\documentclass[10pt,letterpaper,twocolumn]{article} 

\usepackage{ol2}
\usepackage[draft]{hyperref}
\usepackage{amsmath}

\begin{document}

\twocolumn[ 

\title{Frequency-doubled scattering of symmetry-breaking surface-state electrons on liquid Helium}


\author{Miao Zhang,$^{1}$ Wenzhi Jia,$^1$ and Lianfu Wei$^{1,2*}$}

\address{
$^1$Department of Physics, Southwest Jiaotong University, Chengdu
610031, China
\\
$^2$State Key Laboratory of Optoelectronic Materials and
Technologies, \\ School of Physics and Engineering, Sun Yat-sen
University, Guangzhou 510275, China \\
$^*$Corresponding author: weilianfu@gmail.com }

\begin{abstract}Any systems with symmetry-breaking eigenstates can effectively
radiate photons with doubled frequency of the incident light, which
is known as the second harmonic generation. Here, we study the
second-order nonlinear effects with the system of surface-state
electrons on liquid Helium. Due to the symmetry-breaking
eigenstates, we show that a {\it Rabi oscillation} between two
levels of the surface-state electrons can be realized beyond the
usual resonant driving. Consequently, an electromagnetic field with
the doubled frequency of the applied driving could be effectively
radiated. This can be regarded as a frequency-doubled fluorescence,
and interestingly, it works in the unusual Terahertz range.
\end{abstract}

\ocis{270.4180, 300.2530.}

 ] 

{\it Introduction.---} An electron (with mass $m_e$ and charge $e$)
near the surface of liquid Helium is weakly attracted by its
dielectric image potential $V(z)=-\Lambda e^2/z$. Where $z$ is the
distance above liquid Helium surface, and
$\Lambda=(\varepsilon-1)/4(\varepsilon+1)$ with $\varepsilon$ being
dielectric constant of the liquid Helium~\cite{Girmes1}. Due to the
Pauli exclusion principle, there is an barrier about $1$~eV for
preventing the electron penetrating into liquid Helium. Together
with such a hard wall at $z=0$, the electron is resulted in an
one-dimensional (1D) hydrogenlike spectrum $E_n=-R/n^2$ of motion
normal to the Helium surface, with Rydberg energy
$R=\Lambda^2e^4m_e/(2\hbar^2)\approx8$~K and effective Bohr radius
$r_b=\hbar^2/(m_ee^2\Lambda)\approx76$~{\AA}.
These Rydberg levels were first observed by Grimes {\it et
al}.~\cite{Girmes1,Girmes2} by measuring the resonant frequencies of
the transitions between the ground and excited states. Recently, it
has been studied that the above Rydberg states have relatively-long
lifetimes~\cite{saturation}, and possibly, the electrons could be
used as qubits (with the ground and first excited states) to realize
the relevant quantum information
processing~\cite{Science,PRB,experiment-L, experiment-A,
Spin,Simulator,electronQED}.

In this letter, we first study the second-order nonlinear optical
effect of the surface-state electrons on liquid Helium. Due to the
symmetry-breaking surface-states, the electrons allow an unusual
dipole moment which is related to the nonzero average distances
between the Rydberg states, i.e., $d=\langle m|z|m\rangle-\langle
n|z|n\rangle\neq 0$. Basically, under the resonant driving the usual
transition $\langle n |z|m\rangle$ is dominant and the effects
related to $d$ are negligible. Alternatively, under the certain
large-detuning driving, the effects related to $d$ become important,
and consequently a Rabi oscillation between two selected levels of
the surface-state electrons could be realized beyond the usual
resonant driving. This Rabi oscillation leads an electromagnetic
wave emission with the doubled frequency of the applied driving.

{\it Rabi oscillations beyond the resonant driving.---}We consider
applying a microwave field $E_l=\mathcal{E}\cos(\omega_lt-kr+\phi)$
to a single surface-state electron, where $\mathcal{E}$, $\omega_l$,
$k$, and $\phi$ are its amplitude (in $z$ direction), frequency,
wave vector, and initial phase, respectively. For simplicity, we
consider only two levels (i.e., the ground state $|1\rangle$ and the
first excited state $|2\rangle$) of the surface-state electron.
Therefore, under the usual dipole approximation ($kr\approx0$) the
Hamiltonian describing the driven two-level electron reads
$\hat{H}(t)=E_1|1\rangle\langle1|+E_2|2\rangle\langle
2|-e\hat{z}\mathcal{E}\cos(\omega_lt+\phi)$. By using the
completeness relation $|1\rangle\langle
1|+|2\rangle\langle2|=\hat{\textbf{1}}$, one can write $\hat{z}$ as
$\hat{z}=z_{11}|1\rangle\langle 1|+z_{22}|2\rangle\langle
2|+z_{12}|1\rangle\langle 2|+z_{21}|2\rangle\langle 1|$ with the
dipole matrix elements $z_{12}=\langle 1|\hat{z}|2\rangle=\langle
2|\hat{z}|1\rangle=z_{21}$. Due to the asymmetry eigenstates
$|1\rangle$ and $|2\rangle$, here $z_{11}=\langle
1|\hat{z}|1\rangle$ and $z_{22}=\langle 2|\hat{z}|2\rangle\neq0$.
This is very different from the case for the natural atoms wherein
$z_{11}=z_{22}=0$.

Obviously, the Hamiltonian of the electron can be rewritten as
\begin{equation}
\hat{H}(t)=\frac{\hbar\omega_e}{2}\hat{\sigma}_z-\frac{1}{2}\left[\hbar\widetilde{\Omega}\hat{\sigma}_z
+\hbar\Omega_{\text{R}}(\hat{\sigma}_{12}+\hat{\sigma}_{21})\right]
\cos(\omega_lt+\phi),
\end{equation}
and further
\begin{equation}
\begin{array}{l}
\hat{H}(t)=-\hbar\widetilde{\Omega}\hat{\sigma}_z
\left(e^{i\omega_lt+i\phi}+\text{H.c}\right)\\
\\
\,\,\,\,\,\,\,\,\,\,\,\,\,\,\,\,\,\,\,\,
-\hbar\Omega_{\text{R}}\left[\hat{\sigma}_{12}(e^{-i\Delta
t+i\phi}+e^{-i(\omega_e+\omega_l)t-i\phi})+\text{H.c.}\right],
\end{array}
\end{equation}
in the interaction picture defined by
$\hat{R}=\exp(-it\omega_e\hat{\sigma}_z/2)$.
Above, $\hat{\sigma}_z=|2\rangle\langle2|-|1\rangle\langle1|$ is the
Pauli operator with the transition frequency
$\omega_e=(E_2-E_1)/\hbar$, $\hat{\sigma}_{ij}=|i\rangle\langle j|$
is the projection operator (with $i,\,j=1,\,2$),
$\Delta=\omega_e-\omega_l$ is the detuning between the microwave and
electron, $\Omega_{\text{R}}=z_{12}e\mathcal{E}/(2\hbar)$ is the
usual Rabi frequency, and finally
$\widetilde{\Omega}=(z_{22}-z_{11})e\mathcal{E}/(4\hbar)\neq0$ due
to the symmetry-breaking in the states $|1\rangle$ and $|2\rangle$.

Near the resonant point, i.e., $\Delta\ll(\omega_l,\,\omega_e)$, the
evolution ruled by the Hamiltonian (2) can be approximately
expressed as
$\hat{U}(t)=1+\left(\frac{-i}{\hbar}\right)\int_0^t\hat{H}(t_1)dt_1
+\cdots \approx \hat{T}\exp[\int_0^t\hat{H}_{\text{R}}(t)dt]$ with
the Dyson-series operator $\hat{T}$ and effective Hamiltonian
$\hat{H}_{\text{R}}(t)=-\hbar\Omega_{\text{R}}(e^{-i\Delta
t+i\phi}\hat{\sigma}_{12}+\text{H.c.})$.
In the rotating-framework defined by the unitary operator
$\hat{R}_{\text{R}}=\exp(it\Delta\hat{\sigma}_z/2)$, this effective
Hamiltonian reads
\begin{equation}
\hat{H}_{\text{R}}=\frac{\hbar\Delta}{2}\hat{\sigma_z}
-\hbar\Omega_{\text{R}}(e^{i\phi}\hat{\sigma}_{12}
+e^{-i\phi}\hat{\sigma}_{21}).
\end{equation}
Certainly, when $\Delta=0$, the Hamiltonian (3) describes the
standard Rabi oscillation with the frequency $\Omega_{\text{R}}$.
Above, the terms related to
$\xi=(\Omega_{\text{R}},\,\widetilde{\Omega})/(\omega_l,\omega_e)$
are neglected, since $\xi\ll1$ under the weak drivings. In fact,
this approximation is nothing but the usual rotating-wave
approximation. Indeed, under the resonant excitation, i.e.,
$\Delta\ll(\omega_l,\,\omega_e)$, the contribution from
$\hbar\widetilde{\Omega}\hat{\sigma}_z$ (due to the
symmetry-breaking of the present surface-state electrons) should be
negligible, due to its small probability proportional to $\xi$.

Interestingly, under the large-detuning driving, e.g.,
$\Delta=\omega_l-\delta$ (with a small adjustment $\delta\sim0$),
the effects related to the Stark term
$\hbar\widetilde{\Omega}\hat{\sigma}_z$ become significant. This can
be seen from the following new evolution operator~\cite{Nature}
\begin{equation}
\begin{array}{l}
\hat{U}(t)= 1+\left(\frac{-i}{\hbar}\right) \int_0^t\hat{H}(t_1)dt_1
\\
\\
\,\,\,\,\,\,\,\,\,\,\,\,\,\,\,\,\,\,\,\,
+\left(\frac{-i}{\hbar}\right)^2
\int_0^t\hat{H}(t_1)\int_0^{t_1}\hat{H}(t_2)dt_2dt_1 +\cdots\\
\\
\,\,\,\,\,\,\,\,\,\,\,\,\,\,\approx
\hat{T}\exp\left[\int_0^t\hat{H}_{\text{L}}(t)dt\right],
\end{array}
\end{equation}
with the second-order effective Hamiltonian $\hat{H}_{\text{L}}(t)=
\hbar\nu\hat{\sigma}_z/2-\hbar\Omega_\text{L}(e^{i\delta
t+i2\phi}\hat{\sigma}_{12}+\text{H.c.})$, which can be further
written as
\begin{equation}
\hat{H}_{\text{L}}=\frac{\hbar\Delta'}{2} \hat{\sigma}_z -\hbar
\Omega_{\text{L}}(e^{i2\phi}\hat{\sigma}_{12}+
e^{-i2\phi}\hat{\sigma}_{21})
\end{equation}
in the rotating framework defined by
$\hat{R}_{\text{L}}=\exp(-it\delta\hat{\sigma}_z/2)$. Obviously, if
the ``new detuning" $\Delta'=0$, i.e., $\omega_l\approx \omega_e/2$,
the Hamiltonian (5) also defines to a Rabi oscillation with the
frequency $\Omega_{\text{L}}$.

Above, the value of $\delta=2\omega_l-\omega_e$ is controllable by
the selected frequency $\omega_l$ of the applied microwave, and
$\Delta'=\nu-\delta$ with
$\nu=4\Omega_{\text{R}}^2[1/(\omega_e-\delta)+1/(3\omega_e+\delta)]$.
Due to the weak driving
$(\Omega_{\text{R}},\,\widetilde{\Omega})\ll(\omega_l,\omega_e)$,
the present Rabi frequency
$\Omega_{\text{L}}=4\Omega_{\text{R}}\widetilde{\Omega}\omega_e/(\omega_e^2-\delta^2)$
is smaller than the previous one $\Omega_{\text{R}}$ for the
resonant excitation.
Also, all the effects related to the small quantity
$\xi=(\Omega_{\text{R}},\,\widetilde{\Omega})/(\omega_l,\omega_e)$
were neglected, but the terms containing
$\Omega_{\text{R}}\widetilde{\Omega}/(\omega_l,\,\omega_e)$ were
retained.
Note that, under the above large-detuning driving, the usual
first-order expansion term
$\hat{U}^{(1)}(t)=(-i/\hbar)\int_0^t\hat{H}(t_1)dt_1$ practically
does not contribute to the time evolution, due to its small
probability: $P^{(1)}(t)\sim\xi$.

\begin{figure}[htb]
\centerline{\includegraphics[width=7cm]{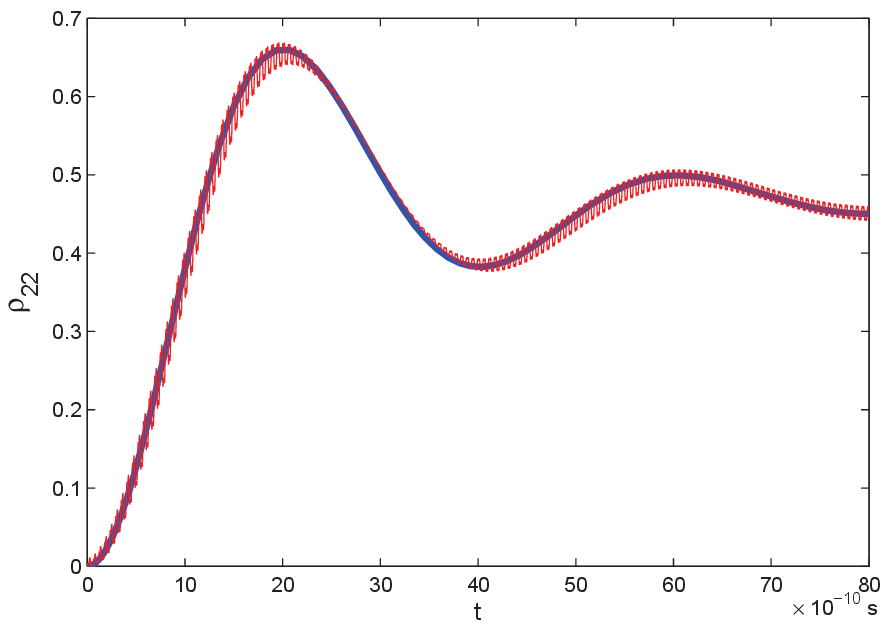}}
\caption{(Color online) Time dependent $\rho_{22}$, with $\Delta'=0$
and $\Gamma=\gamma=g/10$. The blue and red lines correspond to the
analytical solution from the effective Hamiltonian (5) and the
numerical one from the original Hamiltonian (1), respectively. The
relevant parameters are selected as~\cite{Girmes2,PRB}:
$\omega_e\approx0.22$~THz, $z_{12}\approx0.5\,r_b$,
$z_{22}-z_{11}\approx2.3\,r_b$, and $\mathcal{E}\approx15$~V/cm, and
consequently $\Omega_{\text{R}}\approx4.3$~GHz and
$\Omega_{\text{L}}\approx0.8$~GHz.}
\end{figure}

{\it Frequency-doubling radiations.---}We now consider the
steady-state radiations due to the above oscillations of occupancies
by taking into account the practically-existing dissipations. For
this case the dynamics of the driven surface-state electrons is
described by the master equation
\begin{equation}
\begin{array}{l}
\frac{d\hat{\rho}}{dt}=\frac{-i}{\hbar}[\hat{H}_{\text{L}},\hat{\rho}]+
\frac{\Gamma}{2}\left(2\hat{\sigma}_{12}
\hat{\rho}\hat{\sigma}_{21}-\hat{\sigma}_{21}
\hat{\sigma}_{12}\hat{\rho}-\hat{\rho}
\hat{\sigma}_{21}\hat{\sigma}_{12}\right)
\\
\\
\,\,\,\,\,\,\,\,\,\,\,\,\,\,+
\frac{\gamma}{2}\left(2\hat{\sigma}_{22}\hat{\rho}\hat{\sigma}_{22}-\hat{\sigma}_{22}\hat{\rho}-\hat{\rho}
\hat{\sigma}_{22}\right)
\end{array}
\end{equation}
with $\Gamma$ and $\gamma$ being the decay and dephasing rates,
respectively. $\hat{\rho}=\sum_{i,j=1}^{2}\rho_{ij}|i\rangle\langle
j|$ is the density operator of the two-level quantum system, and the
matrix elements $\{\rho_{ij}\}$ obey the normalized and hermitian
conditions: $\sum_{i=1}^{2}\rho_{ii}=1$ and $\rho_{ij}=\rho_{ji}^*$,
respectively. Certainly, due to the resistance from the
surroundings, the amplitudes of oscillating $\rho_{ii}$ decrease
with the time evolving~(see, e.g., Fig. 1), and
$\dot{\rho}_{ii}\rightarrow0$ when $t\rightarrow\infty$.
Specially, under the steady-state condition: $\dot{\rho}_{ij}=0$, we
have $\rho_{22}=2\Omega_{\text{L}}^2\kappa/
[\Gamma(\kappa^2+\Delta'^2)+4\Omega_{\text{L}}^2\kappa]$ and
$\rho_{21}=\Omega_{\text{L}}e^{-i2\phi}
(\rho_{11}-\rho_{22})(i\kappa+\Delta')/(\kappa^2+\Delta'^2)$
with $\kappa=(\Gamma+\gamma)/2$.

Immediately, we can compute the polarization $P=\langle
ez\rangle=\text{Tr}(e\hat{\rho}\hat{z})=e\sum_{i,j=1}^2\rho_{ij}z_{ji}$
for single surface-state electrons. Note that the Eq. (5) is
obtained in the rotating-framework with frequency
$\omega_e+\delta=2\omega_l$. Transforming back to the
Schr$\ddot{\text{o}}$dinger picture, the polarization of the
electrons reads
\begin{equation}
P=e\left(\rho_{21}z_{12}e^{-i2\omega_lt}+\text{H.c.}\right)
+e\left(\rho_{11}z_{11}+\rho_{22}z_{22}\right).
\end{equation}
This indicates that the surface-state electron has a steady
$2\omega_l$-response for the applied incident field of frequency
$\omega_l$. Such a $2\omega_l$-response acts further as an effective
source of new radiation with doubled-frequency $2\omega_l$. By
discarding the static polarizations $\rho_{11}z_{11}$ and
$\rho_{22}z_{22}$ (which practically does not contribute to the
generations of the radiations), we can write the polarization~(7) in
a simple form $P=A\cos\left(2\omega_lt-\theta\right)$
with the amplitude $A=ez_{12}\sqrt{\rho_{21}\rho_{12}}$ and phase
$\theta=\arctan(\text{Im}\rho_{21}/\text{Re}\rho_{21})+2\phi$.

\begin{figure}[htb]
\centerline{\includegraphics[width=7cm]{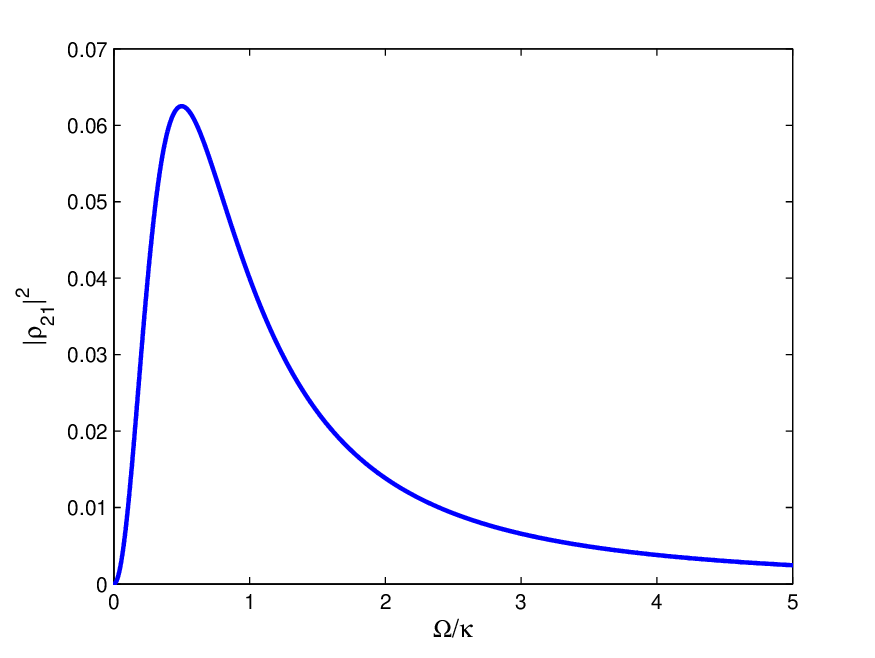}}
\caption{(Color online) Scattering saturation: $|\rho_{21}|^2$
versus $\Omega/\kappa$ with typical $\kappa=\Gamma$ and
$\mathcal{D}=0$. If $\Omega=\Omega_{\text{R}}$ and
$\mathcal{D}=\Delta$ this figure corresponds to that of the usual
resonant scattering, and if $\Omega=\Omega_{\text{L}}$ and
$\mathcal{D}=\Delta'$ the figure corresponds to that of the present
frequency-doubled scattering.}
\end{figure}

Obviously, the radiating power of this electric-dipole reads
$I=A^2(2\omega_l)^4/(12\pi\epsilon_0c^3)$, which is proportional to
$|\rho_{21}|^2$. In the weak field regime, i.e., $\Omega_{\rm
L}\ll\kappa$, the microwave scattering can be significantly enhanced
by increasing $\Omega_{\rm L}$ as shown in Fig.~2. However, if
$\Omega_{\rm L}$ is further enlarged, the scattering appears
saturation and begins to decrease. Note that the Hamiltonian (3)
(for resonant exciting) and the Hamiltonian (5) (for large-detuning
exciting) have the same form, and thus the present frequency-doubled
scattering has the same peak value to that of the usual resonant
scattering.

{\it Discussions and Conclusions.---}As the frequency of microwaves
is relatively-low (compared to the usual optical frequency), the
resonant or frequency-doubled scattering effect of the single
surface-state electron is very weak, e.g., the induced
$I\approx10^{-25}$~W for the typical
$\omega_l=0.22$~THz~\cite{Girmes2}, $z_{12}\approx0.5\,r_b$, and
$|\rho_{21}|^2=0.063$. To significantly enhance the scattering power
an effective approach is to enlarge the transition frequency
$\omega_e$ (since $I\propto\omega_e^4$). For example, if the
transition frequency is controlled as $\omega_e=2.7$~THz, then the
emission power of a single electron can reach
$I\approx2.2\times10^{-21}$~W (i.e., about one photons per second).
This enhanced frequency can be achieved by applying a vertical
electrostatic field $E_\perp\approx0.91$~kV/cm to the
electron~\cite{PRB}. Of course, the scattered power can also be
enhanced by increasing the number of electrons. For the
multi-electrons system, we assume that: (1) the microwave-absorption
is sufficiently small such that the amplitude of the applied
microwave can be regarded unchanged during microwave-electron
interaction, (2) the electron-electron Coulomb
interaction~\cite{Konstantinov} and the sub-drivings of the
scattering waves are negligible, i.e, they are far weaker than the
driving of the applied microwave and the decay of electrons.
Therefore, the total scattering power of $n$ electrons can be simply
expressed as $I_t\approx nI$. In the experiment~\cite{saturation},
the electrons are confined in an area
$s\approx5\,\text{cm}\times5\,\text{cm}$ with a low electron-density
of $\eta\approx0.18\times10^{11}\,\text{m}^{-2}$. With these typical
parameters, we have $I_t\approx s\eta I\approx0.1$~pW for
$\omega_e=2.7$~THz, which could be observed by the THz detectors,
see, e.g.,~\cite{NatureTHz,IEEE-THz}.

In conclusion, we have shown that an effective Rabi oscillation of
the surface-state electrons on liquid Helium can be implemented
under the large-detuning driving, i.e., $\omega_l\approx\omega_e/2$.
Based this Rabi oscillation, we have discussed the polarization (and
consequently the frequency-doubled radiation) according to the
steady-state solution of the relevant master equation. It has been
shown that the present doubled-scattering has the same peak power to
that of the usual resonant scattering. Experimentally, the scattered
power of the single electron is very weak because of the lower
microwave frequency. However, by enlarging the transition frequency
and increasing the number of electrons the scattering power can be
significantly enhanced and could be detected experimentally.

This work was supported by the National Natural Science Foundation
of China Grants No. 11204249, 11147116, 11174373, and 90921010, the
Major State Basic Research Development Program of China Grant No.
2010CB923104.

\end{document}